\documentclass[aps,prl,twocolumn,showpacs]{revtex4} 
\usepackage{epsfig} 
\usepackage{amsmath} 
\usepackage{bbm} 
 
\begin{document} 
 
\title{Probe-configuration dependent  
dephasing in a mesoscopic interferometer} 
 
\author{G. Seelig, S. Pilgram, A.~N. Jordan, and M. B\"uttiker} 
 
\affiliation{ D\'epartement de Physique Th\'eorique, 
 Universit\'e de Gen\`eve, 
CH-1211 Gen\`eve 4, Switzerland} 
 
\date{April 1, 2003}

\begin{abstract} 
Dephasing in a ballistic four-terminal Aharonov-Bohm geometry 
due to charge and voltage fluctuations is 
investigated. 
Treating two terminals as voltage probes, we find a strong dependence
of  the dephasing 
rate on the probe configuration in agreement with a recent experiment by  
Kobayashi et al. (J. Phys. Soc. Jpn. {\bf 71}, 2094 (2002)). 
Voltage fluctuations in the measurement circuit
are shown to be the source of the configuration dependence.

\end{abstract} 
 
 
\maketitle 
 
Recently, Kobayashi et al.\ \cite{kobayashi} measured the reduction
of the Aharonov-Bohm (AB) effect \cite{ab} in a ballistic four-terminal ring
due to decoherence and thermal averaging. 
Not only was the visibility of the AB-oscillations found to be much larger 
in the non-local configuration (see Fig.(1)), but also 
decoherence was observed to be considerably weaker than  
in the local configuration (see Fig.(1)). 
That the external measurement circuit  can strongly influence 
the physical properties of a mesoscopic conductor  
has been shown for a variety of problems ranging from dephasing in disordered 
conductors \cite{aak} to Coulomb blockade \cite{devoret} or 
 the higher moments of the noise in a tunnel  junction \cite{beenakker}.
 However, to our knowledge the experiment of 
Kobayashi et al.~\cite{kobayashi} provides  the first experimental 
evidence of such a striking dependence of the coherence properties  
of open mesoscopic conductors on the {\it measurement configuration}.  
The purpose of this letter 
is to provide a theoretical  explanation of this phenomenon. 
 
In the experiment of Ref.~\cite{kobayashi}, the decoherence rate was 
extracted from a measurement of the four-terminal resistance 
$R_{\alpha \beta,\gamma \delta}$. 
The two contacts $\alpha, \beta$ are voltage biased and monitored by an 
ampmeter while the two contacts $\gamma, \delta$ 
are connected to a voltmeter. In mesoscopic transport, the four-probe  
character of resistance measurements \cite{markus_four_term}  
becomes apparent if the probes  
are within a coherence volume of the sample \cite{webb}. A resistance measurement  
is termed {\it local} if the voltage probes are along the current path  
and is termed {\it non-local} if the voltage probes are far from the current path.  
For the conductor shown in Fig.~(1), $R_{14,23}$ is a local resistance, whereas  
$R_{12,34}$ is an example of a non-local resistance.  
We emphasize that the sample is the same, independent of the resistance  
measured: what changes is how the sample is connected to the current source  
and to the voltmeter.  

AB oscillations are the result of
quantum interference from electrons travelling through the two arms of
the ring. In ballistic mesoscopic rings these oscillations can be larger than 50\% of the total
current amplitude \cite{kobayashi,kayser,ji},
 and their decay is a measure of decoherence in
the system (once thermal averaging is taken into account).
Experimental investigations \cite{casse,hansen} found a linear temperature
 dependence of the dephasing rate \cite{hansen}.
A theoretical explanation, starting from
fluctuating electrostatic potentials in the ring, is given in Ref.~\cite{seelig}.
Similar results for the temperature dependence of the dephasing rate, both experimental \cite{bird,clarke} and theoretical \cite{takane}, have been obtained previously for chaotic quantum dots. 
Here, we are concerned with another feature of the dephasing rate,
namely its probe configuration dependence \cite{kobayashi}.

\begin{figure} 
\label{koba_fig} 
\begin{center} 
\epsfig{file=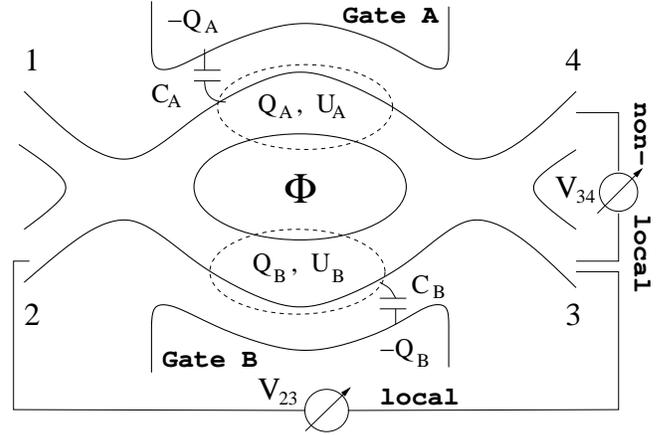, width=8.5cm} 
\caption{Ballistic four-terminal ring:  An internal potential $U_i(t)$  
and a charge  $+Q_i(t)$ belong to each arm of the ring  ($i=A,B$).
 Each arm is coupled to a side-gate via a capacitance $C_i$.
The local and non-local voltage-probe configurations are indicated.} 
\end{center} 
\vspace{-.5cm}
\end{figure} 

First, we illustrate our approach for a (reflectionless) electronic Mach-Zehnder  
interferometer (MZI) \cite{seelig}. In a second step, we consider  interferometers 
with backscattering at the intersections. In both cases, 
the arms of the ring are treated as perfect one-channel leads 
 that can be charged up relative to nearby side-gates via the capacitances $C_A$
and $C_B$. The setup is sketched in Fig.~(1). 
For the MZI, the intersections are described as reflectionless beam splitters
 (see inset in Fig.~(2))  with a scattering matrix 
\begin{equation}\label{beamsplitter} 
{\bf S}_B=\left(\begin{array}{ll} 
           0&{\bf s}\\ 
           {\bf s}&0 
          \end{array}\right),\,\,\,\, 
{\bf s}=\left(\begin{array}{cc} 
           \sqrt{{\cal T}}&i\sqrt{1-{\cal T}}\\ 
           i\sqrt{1-{\cal T}}&\sqrt{{\cal T}} 
          \end{array}\right). 
\end{equation} 
Here, $\sqrt{{\cal T}}$ is the amplitude for going straight through the intersection  
 and $i\sqrt{1-{\cal T}}$ the amplitude for being deflected. 
Due to the absence of backscattering, the MZI does not exhibit 
closed electronic trajectories.

Electron-electron interactions give rise to fluctuations of the internal 
potentials $U_i(t)$ ($i=A,B$). In the presence of interactions, the 
 dimensionless conductance  $G_{13}$ relating current at contact 1 to a voltage 
 applied to contact 3 is \cite{seelig} 
\begin{equation}\label{conductance} 
G_{13}=-\langle T_{13}\rangle =-2{\cal T}(1-{\cal T})\left[1+{\rm e}^{-\tau\Gamma_\phi}
\cos \left( 2\pi\Phi \right)\right]. 
\end{equation} 
Here, $T_{13}$ is the transmission probability,
$\tau=L/v_F$ is the traversal time, $\Phi$ is the 
magnetic flux through the ring (in units of the  
flux quantum) and we have taken equal arm lengths in the ring,
 $L_A=L_B=L$. 
The angular brackets  denote an  average
over the potential fluctuations in the ring  \cite{thermal}.
In the limit of classical Nyquist noise, the decoherence rate 
\begin{equation}\label{dephasing_spectrum} 
\Gamma_\phi=\frac{e^2}{2\hbar^2}\Sigma_{UU}(0) 
\end{equation} 
is then proportional to the spectral function $\Sigma_{UU}(0)$ 
of the potential difference $U(t)=U_A(t)-U_B(t)$ in the zero-frequency limit. 
If all four contacts are connected to a zero-impedance external circuit
 kept at constant voltage, the rate $\Gamma_\phi$ of dephasing due to  
(small energy transfer) electron-electron scattering is   
\begin{equation}\label{rate0} 
\gamma^0_\phi=\frac{2kTe^2}{\hbar^2}\left(\frac{C_\mu}{C}\right)^2R_q. 
\end{equation} 
Here, $T$ is the temperature, $R_q=h/(4e^2)$ is the charge 
relaxation resistance and 
 the electrochemical capacitance $C^{-1}_\mu=C^{-1}+(e^2D)^{-1}$  
 is the series combination of the geometrical  capacitance and the  
density of states \cite{ap1}. We assumed $C_A=C_B=C$.
 
In the experiment of Ref.~\cite{kobayashi}, two probes 
are connected to a voltmeter which ideally has infinite impedance.  
The voltage at a lead connected to the voltmeter  
fluctuates to maintain zero net current.  
These voltage fluctuations give rise to fluctuations of the internal 
potentials which in turn leads to additional dephasing.  For the interferometer 
shown in Fig.~(1), this new contribution to the dephasing rate turns out to  
depend strongly on the probe configuration. For the dephasing rates 
in the local ($\it l$) and non-local ($\it nl$) configuration we obtain
 respectively, 
\begin{subequations}
\label{gammas}
\begin{eqnarray} 
\Gamma^{l}_\phi=\gamma^0_\phi+\gamma^{l}_\phi,&&\,\,\,\gamma^{l}_\phi=\gamma^0_\phi\,\frac{(2{\cal T}-1)^2}{2{\cal T}(1-{\cal T})+T_0},\label{rate_local}\\ 
\Gamma^{nl}_\phi=\gamma^0_\phi+\gamma^{nl}_\phi,&&\,\,\,\gamma^{nl}_\phi=\gamma^0_\phi\,\frac{(2{\cal
    T}-1)^2}{1+2T_0}. \label{rate_nonlocal} 
\end{eqnarray}
\end{subequations}
Here, $\gamma^{l}_\phi$ and $\gamma^{nl}_\phi$ are  the probe-configuration 
specific contributions.  The experiment of Ref.~\cite{kobayashi} shows 
 transmission between neighboring contacts to be significant.
For better comparison, we therefore included a finite incoherent transmission
 $T_0=T_{12}=T_{21}=T_{34}=T_{43}$ (see Fig.~(2)).

The  results for the  dephasing rates are strongly 
 dependent on the symmetry of the interferometer.  
In the symmetric case (${\cal T}=1/2$) the contribution 
to the dephasing rate due to  voltage fluctuations  vanishes for  
both measurement configurations.  
\begin{figure} 
\label{rate_fig} 
\begin{center} 
\epsfig{file=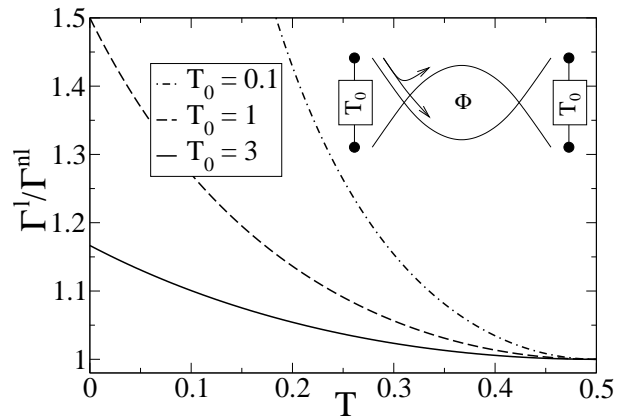, width=8cm} 
\caption{The ratio of the local to the non-local decoherence rate 
is shown as a function of the  
transmission ${\cal T}$ at the beam splitters for different values of the incoherent parallel resistance $1/T_0$. 
All curves are symmetric with respect to ${\cal T}=1/2$. 
 In the inset, the two possible electron paths at the beam-splitter 
 and the  resistance $1/T_0$ 
(in units of $h/e^2$) are indicated.} 
\end{center} 
\vspace{-.5cm}
\end{figure}  
Away from the symmetry point, ${\cal T}=1/2$, the local 
and non-local decoherence rate differ strongly \cite{pilgram}.
A local decoherence rate several times larger than the non-local
one can easily be obtained for small enough $T_0$. 
The ratio of the decoherence rates for the two probe configurations is 
shown in  Fig.~(2) as a function of the transmission  probability ${\cal T}$.  

To derive the results presented in Eqs.~(\ref{rate_local}) and 
(\ref{rate_nonlocal}),
we need to know the spectral function  $\Sigma_{UU}(0)$ 
for the two  different probe configurations.  
To start with, we want to express  
 the Fourier transform $\Delta U(\omega)=\Delta U_A(\omega)-\Delta U_B(\omega)$ 
 of the fluctuations of the internal potential operator through the operators for  
the bare charge $\Delta Q^b_i(\omega)$ ($i=A,B$) and current $\Delta I^b_{\alpha}(\omega)$
($\alpha=1,\ldots,4$) fluctuations in the sample. For these quantities  it is known how to calculate 
the spectral functions. 
The notation $\Delta{\cal O}={\cal O}-\langle{\cal O}\rangle$ 
 denotes deviations of an operator $\cal O$ from its expectation value. 
There are two independent equations relating charge and potential fluctuations, namely, 
\begin{equation}\label{charge} 
\Delta Q_i=C\Delta U_i=\Delta Q^b_i-e^2D \Delta U_i+e^2\sum_{\alpha} D^{(i)}_{\alpha}\Delta V_\alpha. 
\end{equation} 
Here, $\Delta Q^b_i(\omega)$ are the   
charge fluctuations at constant voltage and internal potential
and $\Delta V_\alpha$ are the voltage fluctuations at contact $\alpha$. 
The response to a change in the applied voltage at contact $\alpha$ 
is determined by the average injectivity   
$D^{(i)}_{\alpha}(\omega)=\langle D^{(i)}_{\alpha\alpha}(\omega)\rangle=\partial \langle Q_i\rangle /e^2\partial V_\alpha$. 
 The term with the negative sign in Eq.~(\ref{charge}) is the screening charge induced in response 
 to a change in the internal potential. In the Thomas-Fermi approximation,  
  the response function is the  density  
$D(\omega)=-\partial \langle Q_i \rangle /e^2\partial U_i$. 
 For zero frequency, we find $D=\sum_{\alpha}D^{(i)}_{\alpha}=2L/hv_F$
as a consequence of gauge invariance.  
The injectivities are the diagonal elements of the local density of states (DOS) 
 matrix \cite{pedersen,martin} which is related to the 
scattering matrix $S_{\alpha\beta}$ of the system. 
In the zero-frequency limit 
\begin{equation}\label{lDOS} 
D^{(i)}_{\alpha\beta}(E)=-\frac{1}{2\pi i}\sum_\gamma S^\dagger_{\gamma\alpha}(E)\frac{dS_{\gamma\beta}(E)}{edU_i}
\end{equation} 
with $i=A,B$ and $\alpha,\beta,\gamma=1,\ldots,4$.
The scattering matrix for the electronic interferometer  
 can be derived using Eq.~(\ref{beamsplitter}) (see also  Ref.~\cite{seelig}). 
 
 The voltage fluctuations entering Eq.~(\ref{charge}) are related to
 the current fluctuations $\Delta I_\alpha$  through \cite{markus_noise}  
\begin{equation}\label{langevin} 
\Delta I_\alpha=\Delta I^b_{\alpha} +\sum_{\beta} G_{\alpha\beta}\Delta V_\beta. 
\end{equation} 
In this Langevin-like equation, $\Delta I^b_\alpha$ are the bare current fluctuations 
 and $G_{\alpha\beta}=G_0(M_\alpha\delta_{\alpha\beta}- \langle T_{\alpha\beta}\rangle )$
 are elements of the conductance matrix ($\alpha,\beta=1,\ldots,4$). 
For the MZI we have $M_\alpha=1+T_0$  where $T_0$ is the incoherent
 transmission between neighboring external leads  at the same intersection.
 The probabilities for coherent transmission are $T_{\alpha\beta}=|S_{\alpha\beta}|^2$.
  The central new ingredient in this paper are
the boundary conditions  imposed  on the voltage fluctuations 
 $\Delta V_\alpha$ and the fluctuations  
of the total currents $\Delta I_\alpha$ by the external measurement
circuit.
These boundary conditions depend on the measurement configuration.   
In the {\it local configuration} (see Fig.~(1)) we choose contacts $1$ and $4$  
as the current probes  (they exhibit no voltage fluctuations: $\Delta V_1=\Delta V_4 =0$) 
 while contacts  $2$ and $3$ are the voltage probes  
(no current fluctuations: $\Delta I_2=\Delta I_3=0$). 
In the {\it non-local configuration}, on the other hand, 
the voltage probes  are contacts 3 and 4 (cf. Fig.~(1)) and thus 
$\Delta V_1=\Delta V_2=0$ and $\Delta I_3=\Delta I_4=0$.  
Eq.~(\ref{langevin}), together with the boundary conditions for voltages  
and currents,  can now be used to eliminate the voltage fluctuations
in Eq.~(\ref{charge}) 
in favor of the fluctuations of the bare currents. The potential fluctuations $\Delta U$ 
can then be expressed through the fluctuations of 
the bare currents $\Delta I^b_{\alpha}$ and charges $\Delta Q^b_i$. 
The result for $\Delta U$ will be different for the local and non-local  
configuration as a consequence of the different boundary conditions. 
 
The spectral function of the potential fluctuations is  
defined  through the relation 
$ 2\pi\delta (\omega+\omega')\Sigma_{UU}(\omega)= 
\langle \Delta U(\omega)\Delta U(\omega')+\Delta U(\omega') 
\Delta U(\omega)\rangle/2$. Since we now know how to express the potential  
fluctuations for the local and non-local 
case through the fluctuations of the bare currents and charges,  
we can also express the spectral  function  $\Sigma_{UU}(\omega)$,  
through the correlators of the bare charge 
  $\Sigma_{Q^b_iQ^b_k}(\omega)$ ($i,k=A,B$),  the current correlators  $\Sigma_{I^b_{\alpha}I^b_{\beta}}(\omega)$ 
 ($\alpha,\beta=1,\ldots,4$) and  the cross-correlators $\Sigma_{Q^b_iI^b_{\alpha}}(\omega)$ between charges and currents. 
For zero frequency and in the classical limit, the correlator of the charge  
fluctuations $\Delta Q^b_i$ and $\Delta Q^b_k$ in arms $i$ and $k$ is \cite{pedersen,martin} 
\begin{equation}\label{chargenoise1} 
\Sigma_{Q^b_iQ^b_k}(0)=kTh\sum_{\alpha\beta}\langle D^{(i)}_{\alpha\beta}D^{(k)}_{\beta\alpha}\rangle=\delta_{ik}kTDh/2.  
\end{equation} 
The second equation is obtained from Eq.~(\ref{lDOS}) and the 
scattering matrix of the interferometer (see Ref.~\cite{seelig}). 
Finally, the current correlation functions are given by the 
generalized Nyquist formula \cite{markus_noise},  
$\Sigma_{I^b_{\alpha}I^b_{\beta}}(0)=kT\left(G_{\alpha\beta}+G_{\beta\alpha}\right)$, 
while cross-correlations between fluctuations of the bare charge in
arm $k$ and current fluctuations at contact $\alpha$ vanish 
($\Sigma_{Q^b_iI^b_{\alpha}}(0)=0$) because of the absence of backscattering in
our model.
 
We are now in a position to calculate the spectrum of the potential  
fluctuations in the zero-frequency limit. In the local configuration we obtain 
\begin{equation}\label{s_local} 
\Sigma_{UU}(0)=4kTR_q\left(\frac{C_{\mu}}{C}\right)^2\left[1+\frac{(2{\cal T}-1)^2}{T_0+\langle T_{13}\rangle}\right]. 
\end{equation}  
From comparison with Eqs.~(\ref{conductance}) and (\ref{dephasing_spectrum}),
 it becomes clear that Eq.~(\ref{s_local}) is a self-consistent 
equation for the dephasing rate. 
In the limit of weak decoherence, the transmission probability
entering Eq.~(\ref{s_local}) is flux dependent. 
In contrast, for the limit of strong decoherence,  
we can neglect the flux dependence of $\langle T_{13}\rangle $ in Eq.~(\ref{s_local}).   
Using Eq.~(\ref{dephasing_spectrum}), we  then
obtain the local dephasing rate Eq.~(\ref{rate_local}). 
In the non-local case, the spectral function is independent of the 
magnetic field  even when dephasing is weak.
 It is given by  
$\Sigma_{UU}(0)= (2\hbar^2/e^2 )\Gamma^{nl}_\phi$, 
leading to the dephasing rate for the non-local configuration 
 Eq.~(\ref{rate_nonlocal}) \cite{jordan}.

In the MZI, the intersections between contacts and arms  are 
described by ideal beam splitters (see Eq.~(\ref{beamsplitter})).  Backscattering was  
included only through the incoherent transmission $T_0$ between neighboring  
contacts. Ideal beam splitters are rarely realized in an experiment where 
it is probable that scattering in the intersections exhibits a certain degree 
of randomness.  For better comparison with the experimental situation, 
we now investigate numerically a model that interpolates between the ideal beam splitter 
and fully random scattering. The corresponding scattering matrix for one intersection is  
\begin{equation} \label{chaos}
{\bf S}=\sqrt{1-\varepsilon}\,{\bf S}_B-\varepsilon\, {\bf S}_B{\bf S}_C\left[{\mathbbm 1}-\sqrt{1-\varepsilon}{\bf S}_B {\bf S}_C\right]^{-1} {\bf S}_B 
\end{equation} 
where ${\bf S}_B$ is given in Eq.~(\ref{beamsplitter}) and ${\bf S}_C$
 is a random matrix chosen from the circular orthogonal ensemble 
\cite{brower}. 
 The parameter $\varepsilon$ controls the admixture of chaos, 
 $\varepsilon=0$ corresponds to the ideal beam-splitter, while
 $\varepsilon=1$ corresponds to completely random scattering.
\begin{figure} 
\label{chaos_fig} 
\begin{center} 
\epsfig{file=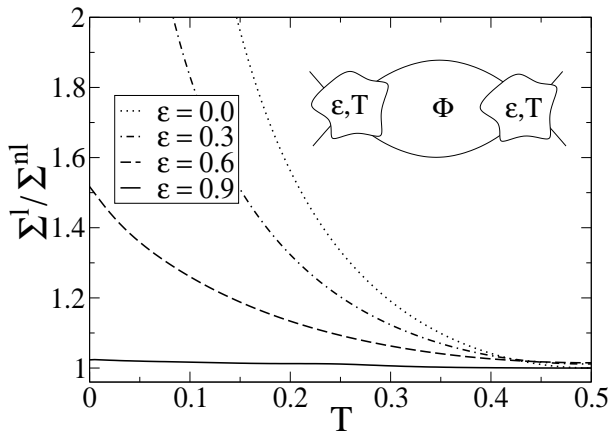, width=8cm} 
\caption{Ratio of the local to the non-local potential fluctuation spectrum
as a function of the transmission probability ${\cal T}$ at the beam splitters
for different values of $\varepsilon$. The parameter $\varepsilon$ controls the 
admixture of chaotic scattering. All curves are
symmetric around ${\cal T}=1/2$.} 
\end{center}
\vspace{-.5cm} 
\end{figure}  
In Fig.~(3), the  ratio of local to non-local potential fluctuations 
 is shown for different values of the parameter $\varepsilon$. 
The results given there are valid in the limit of strong dephasing (as in Fig.~(2)) 
and include an ensemble average over the random matrices
 ${\bf S}_C$ of the two intersections. 
From Fig.~(3) it is clear that increasing the degree of chaotic scattering in the  
intersections suppresses the difference between the local and non-local configuration.  
Backscattering, on the ensemble average, thus has a similar effect as an incoherent parallel transmission 
$T_0$ between neighboring contacts. 
In the limit where the intersections are  
fully chaotic ($\varepsilon = 1$), there is 
no difference between the local and non-local configuration. 
The reason is that ensemble averaging makes the ring 
symmetric to any measurement configuration.

In the experiment of Ref.~\cite{kobayashi}, the four-terminal resistance  
  $R_{\alpha\beta,\gamma\delta}=(V_\gamma-V_\delta)/I_\alpha$ 
with $I_\beta=-I_\alpha$ was measured. In terms of the conductance
 matrix elements, the four-terminal    resistances are  
$ R_{\alpha\beta,\gamma\delta}=\left(G_{\gamma\alpha}G_{\delta\beta}-
G_{\gamma\beta}G_{\delta\alpha}\right)/D$ \cite{markus_four_term}, 
where $D$ is any sub-determinant of rank three of the total
conductance matrix.  The four-terminal resistance takes a particularly simple
form in the case of a reflectionless interferometer ($\varepsilon=0$) where  we find
\begin{subequations}
\label{resistances}
\begin{eqnarray}
R_{14,23}&=&\frac{h}{2e^2}\frac{T_0- \langle T_{13}\rangle}{1+T_0-\langle T_{13}\rangle},\label{four_terminal_l}\\
R_{12,43}&=&\frac{h}{2e^2}\frac{1-2\langle T_{13}\rangle}{(T_0+ \langle T_{13}\rangle)(1+T_0- \langle T_{13}\rangle)}\label{four_terminal_nl}
\end{eqnarray}
\end{subequations}
for the local and non-local configuration, respectively.
Eqs.~(\ref{four_terminal_l}) and (\ref{four_terminal_nl}) show that the attenuation
of the local and non-local resistances is determined by the decoherence
 rates Eqs.~(\ref{rate_local}) and (\ref{rate_nonlocal}) respectively.

In conclusion, 
we have shown that the electrical constraints imposed by the  
measurement circuit give rise to a probe configuration dependence 
of the dephasing rate. This effect is most pronounced in an ideal 
quantum interferometer that is strongly asymmetric, but was found 
to persist even in the presence of a considerable admixture 
of incoherent transmission or randomness (with ensemble averaging).  
While there may be other physical mechanisms for producing such a
difference, our discussion of dephasing  
explicitly includes the effect of the external electrical circuit and 
leads to a result consistent with unanticipated experimental observations. 

This work was supported by the Swiss National Science Foundation.


\begin{thebibliography}{99}
\bibitem{thermal}
AB oscillations in a MZI with two arms of equal length are not affected 
by thermal averaging.
\bibitem{pilgram}
If $T_0=0$, the local dephasing rate Eq.~(\ref{rate_local})  
diverges for ${\cal T}=0$ and ${\cal T}=1$.
The divergence has its origin in the fact that for ${\cal T}=0$ and ${\cal T}=1$
the two arms of the MZI become completely decoupled. It is therefore 
cut off by any backscattering.
\end{thebibliography}
\end{document}